# Toward a Minecraft Mod for Early Detection of Alzheimer's Disease in Young Adults

Satoko Ito
Graduate School of Information
Science and Engineering
Ritsumeikan University
Japan
is0395hr@ed.ritsumei.ac.jp

Ruck Thawonmas
College of Information Science and
Engineering
Ritsumeikan University
Japan
ruck@is.ritsumei.ac.jp

Pujana Paliyawan
Research Organization of Science and
Technology
Ritsumeikan University
Japan
pujana.p@gmail.com

## Abstract

This paper proposes a Minecraft-based system for early detection of Alzheimer's disease in young adults. Early detection, where spatial navigation is a crucial key, is regarded as an important way to prevent the disease. The proposed system is compared with a recent existing and thoroughly studied system using a game called Sea Hero Quest (SHQ), by analyzing spatial navigational patterns of players. Our preliminaries results show that spatial navigational patterns in both systems are highly correlated, indicating that the proposed system is likely as effective as the SHQ system for the detection task.

## 1  Introduction

Dementia is a social problem worldwide, with the expected number of cases to increase to 152 million by 2050 [4]. There exist a number of games developed for early detection of dementia, especially Alzheimer's disease (AD), and among them, we consider the system using Sea Hero Quest (SHQ) [1] to be the most promising candidate, based on their analysis that compared navigational patterns in big data with those from a highly phenotyped healthy aging laboratory cohort.

Our aim is to increase the penetration rate of AD early detection by implementing mechanisms presented in SHQ on a more popular game such as Minecraft. In particular, we aim at detecting young-onset dementia, which can occur as early as from 17 years [3]. This work is a first step toward this aim. We investigate whether navigational results in the proposed Minecraft-based system and the SHQ system are correlated.

## 2  Related Work

SHQ [1] was developed based on the idea that spatial navigation, which can be analyzed from how the player travels in game, is an important key for detection of the preclinical stage of AD. They acquired big data ($n = 27,108$) from players, and through analyses, their results confirmed that SHQ can distinguish high-risk people with AD from healthy elderly people. However, SHQ is a game specially developed for preclinical AD detection, so its penetration rate is likely to be limited when compared with other systems using more popular games such as Minecraft.

Minecraft is not only popular among game players[1], but is also used by many researchers [2, 5]. With total sales of over 200 million, it is still known as the most popular PC game in the world, nearly 10 years after its release. In addition, the availability of "mod", a kind of expansion pack that users can freely develop, allows modification at the game system level. Because Minecraft is a game that is familiar to many people, including young adults, it is highly likely that its mods developed for serious applications such as ours will become widespread.

## 3  Overview of the Proposed Mod

First, when the player enters a world, as shown in Fig. 1.a, a map and a signboard appear, in front of the player, displaying the locations of the starting point and checkpoints, individually associated with a number instructing the visiting order. The signboard shows a list of items that the player must correctly acquire at each checkpoint, one item per checkpoint. Once having grasped the information on where to go and what to acquire, the player then moves to the first checkpoint.

Next, when approaching a checkpoint, the player will see an area with the brick color on the ground, as shown in Fig. 1.b. At each checkpoint are located a chest and a signboard, which will be added by a crafting table at the final checkpoint, as shown from right to left in Fig. 1.c, where the number on the signboard shows the ID of the current checkpoint. When the player opens the chest, a pop-up like the one in Fig. 1.d will appear. The area consisting of the top three rows of this figure shows the contents of the chest, and the bottom four rows construct the so-called inventory for storing acquired items. The player then takes out an item, which the player thinks is a correct one, from the chest and moves it to the inventory.

After items from the chests at all the checkpoints are acquired, craft on the crafting table located at the final checkpoint will be conducted. In this mod, crafting is the creation of a new item using all the acquired items. When the player opens the crafting table, a pop-up like the one in Fig. 1.e will be displayed. In this figure, the upper half is the craft area of the crafting table, while the inventory is displayed in the lower half. After all the acquired items are moved to any of the 3x3 cells in the craft area, a new item will be created on the right side of the arrow. Note that only when a target item is correctly acquired at each checkpoint, a final item will be created; otherwise, a different one. The player then moves the newly created item to any cell in the inventory, after which the current world ends.

---

[1] https://www.statista.com/statistics/680139/minecraft-active-players-worldwide/

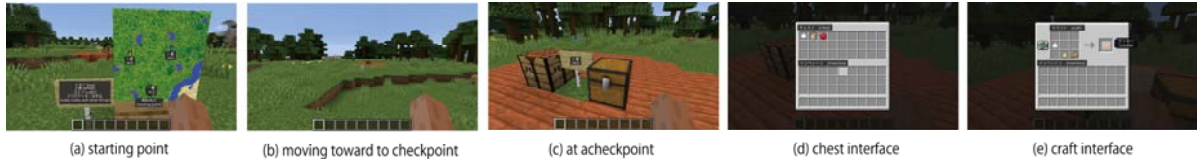

Figure 1: Screenshots of typical gameplay in our Minecraft mod.

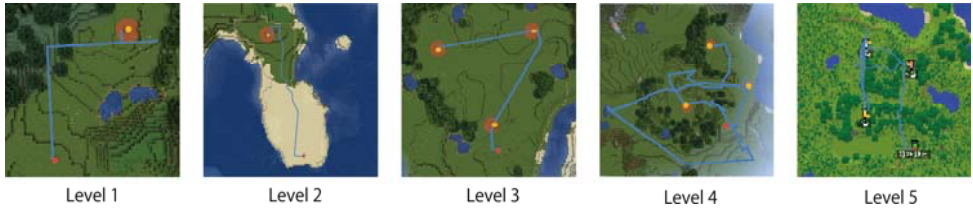

Figure 2: Typical trajectories on each Minecraft level.

## 4 Experiment

In our experiment, there were 10 participants[2], 6 males and 4 females aged 20 to 26 years. They were divided into two groups and asked to play both games on five levels, in increasing order of difficulty, with each group playing SHQ and Minecraft in an opposite order. Each Minecraft level was designed based on its counter-part SHQ level. Please refer to our supplementary page[3] for the layouts of all the levels in both SHQ and Minecraft, player trajectories on them, the source code of the proposed system, and a demo video of our mod's gameplay.

**4.1 Metric for evaluation** To confirm whether our mod has the same spatial navigational patterns as SHQ, the Pearson correlation coefficient ($r$) is derived between the normalized distances, defined in [1], traveled during the play of SHQ and Minecraft.

**4.2 Results and discussions** The Pearson's $r$ of 0.95 is obtained. This indicates a very high correlation [6], and thus supports the possibility to be able to detect people with a high risk of developing preclinical AD, in particular young-onset AD due to the ages of our participants, by using the proposed system. Figure 2 shows play trajectories of a typical participant on all the Minecraft levels in use.

## 5 Conclusions and future work

As a first step toward detection of preclinical Alzheimer's disease in young adults using Minecraft, we proposed a system and compared the distance traveled during its play with the distance traveled during SHQ play. There was a strong correlation between the two distances, which provides an insight that both systems are equally interchangeable. Our future work includes conducting a user study to examine if the proposed system provides better user experience, such as engagement and enjoyment, than the SHQ system; if this is the case, a high penetration rate of the proposed system can be expected.

---

[2] We obtained informed consent, designed based on the research ethics guidelines at the authors' university, from all the participants. According to the guidelines, our research does not require an ethics approval.

[3] https://it0sat0.github.io/mc-mci-supportpage/